\documentclass[12pt]{article} 

\usepackage{latexsym} 
\usepackage{amssymb}  
\usepackage{epsfig}       

\newcommand{\ep}{\varepsilon}

\newcommand{\beq}{\begin{equation}}
\newcommand{\eeq}{\end{equation}}
\newcommand{\ba}{\begin{array}}
\newcommand{\bea}{\begin{eqnarray}}
\newcommand{\ea}{\end{array}}
\newcommand{\eea}{\end{eqnarray}}

\newcommand\comment[1]{ \hbox{[{\it Comment suppressed here.}\/]} }
\newcommand\hide[1]{}

\newcommand{\skipover}[1]{}
\newcommand{\nn}{\nonumber \\}

\makeatletter 

\def\appendix{\par                              
    \setcounter{section}{0}                     
    \setcounter{subsection}{0}
    \renewcommand{\theequation}{\Alph{section}.\arabic{equation}}
    \renewcommand{\thesection}{Appendix \Alph{section}
                \setcounter{equation}{0}  } 
}

\def\applabel#1{\@bsphack
  \protected@write\@auxout{}%
         {\string\newlabel{#1}{{\Alph{section}}{\thepage}}}%
  \@esphack}

\def\section{
\setcounter{equation}{0}        
\@startsection {section}{1}{\z@}{-3.5ex plus -1ex minus 
 -.2ex}{2.3ex plus .2ex}{\large\bf}}
\renewcommand{\theequation}{\arabic{section}.\arabic{equation}}

\def\subsection{\@startsection{subsection}{2}{\z@}{-3.25ex plus -1ex minus 
 -.2ex}{1.5ex plus .2ex}{\normalsize\bf}}

\def\subsubsection{\@startsection{subsubsection}{3}{\z@}{-3.25ex plus
 -1ex minus -.2ex}{1.5ex plus .2ex}{\normalsize}}

\makeatother   

\newsavebox{\eqlabel}

\makeatletter  
\newlength{\numblen}
\newsavebox{\eqnumb}
\def\@eqnnum{\savebox{\eqnumb}{\rm (\theequation)}%
\settowidth{\numblen}{\usebox{\eqnumb}}%
\makebox[\numblen][l]{\usebox{\eqnumb}~~~\usebox{\eqlabel}}}
\makeatother   

\newenvironment{equationwithlabel}[1]{ %
  \savebox{\eqlabel}{#1}
  \begin{equation}\label{#1} }{\end{equation}} 
\newcommand{\beql}[1]{\begin{equationwithlabel}{#1}}
\newcommand{\eeql}{\end{equationwithlabel}}

\begin{document}

\title{\bf Slowing Out of Equilibrium Near the QCD Critical Point}

\author{
	Boris Berdnikov and Krishna Rajagopal \\[0.5ex]
{\normalsize Center for Theoretical Physics}\\
{\normalsize Massachusetts Institute of Technology}\\
{\normalsize Cambridge, MA 02139 }
}

\newcommand{\preprintno}{
  \normalsize MIT-CTP-2931 
}

\date{December 8, 1999 \\[1ex] \preprintno}

\begin{titlepage}
\maketitle
\def\thepage{}          

\begin{abstract}
The QCD phase diagram may feature a critical end point
at a temperature $T$ and baryon chemical potential $\mu$
which is accessible in heavy ion collisions.   
The universal long wavelength fluctuations which
develop near this Ising critical point result in 
experimental signatures which can be used to find the critical point.
The magnitude of the observed effects depends 
on how large the correlation length $\xi$ becomes.  Because
the matter created in a heavy ion collision
cools through the critical region of the phase diagram
in a finite time, critical slowing down limits the growth
of $\xi$, preventing it from staying in equilibrium.
This is the fundamental nonequilibrium effect which
must be calculated in order to make quantitative predictions for
experiment.  We use universal nonequilibrium dynamics and 
phenomenologically motivated values for the necessary
nonuniversal quantities to estimate 
how much the growth of $\xi$ is slowed.
\end{abstract}

\end{titlepage}

\renewcommand{\thepage}{\arabic{page}}


\section{Introduction}
\subsection{The Critical Point}

One goal of relativistic heavy ion collision experiments
is to explore and map the QCD phase diagram as a function
of temperature and baryon chemical potential.
Recent theoretical developments
suggest that a key qualitative feature, namely a critical
point which in a sense defines the landscape
to be mapped, may be within reach of discovery and analysis
by the CERN SPS or by RHIC, if data is taken at several different
energies \cite{SRS1,SRS2}.
The discovery of the critical point
would in a stroke transform the map of the QCD phase
diagram from one based only on
reasonable inference from universality, lattice gauge theory
and models into one with a solid experimental basis \cite{Review}.

In QCD with
two massless quarks ($m_{u,d}=0$; $m_s=\infty$)
the phase transition at which chiral symmetry is restored 
is likely second order and belongs to the universality
class of $O(4)$ spin models in three dimensions \cite{piswil}.
Below $T_c$, chiral symmetry is broken and there are three
massless pions.  At $T=T_c$, there are four massless degrees
of freedom: the pions and the sigma. Above $T=T_c$, the pion
and sigma correlation lengths are degenerate and finite.

In nature, the light quarks are not massless.  Because
of this explicit chiral symmetry breaking,
the second order phase transition is replaced by an 
analytical crossover: physics changes dramatically but smoothly in the 
crossover region, and no correlation length diverges.
This picture is consistent with present lattice 
simulations \cite{latticereview},
which suggest $T_c\sim 140-170$ MeV \cite{latticeTc}.

\begin{figure}[thb]
\begin{center}
\epsfig{file=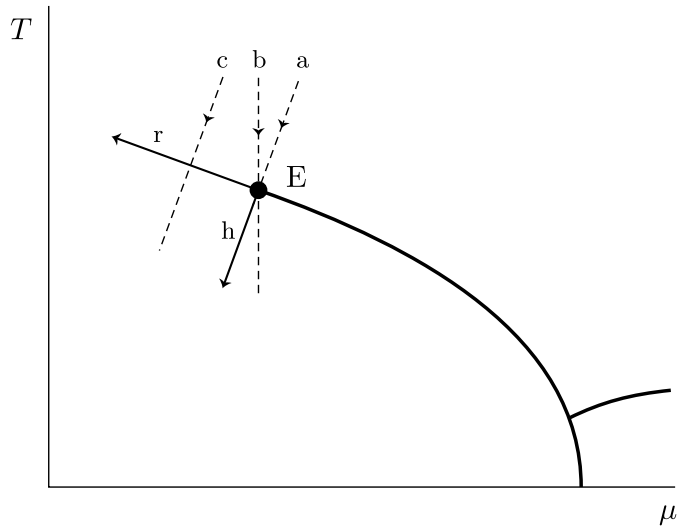,width=3.5in}
\end{center}
\caption{Sketch of the QCD phase diagram as a function of 
temperature $T$ and baryon chemical potential $\mu$.   
Chiral symmetry
is broken at low $T$ and $\mu$.  As $T$ is increased, chiral
symmetry is approximately restored via a smooth crossover to the
left of $E$ or a first order phase transition to the right of $E$.
The symmetry is only approximately restored because the light quarks
are not massless.
At the critical point $E$ at which the line
of first order phase transitions ends, the transition is second order 
and is in the Ising universality class.  
(At large $\mu$
and small $T$, there are color superconducting phases which
we do not discuss in this paper.)
The Ising model $r$-axis and $h$-axis and the trajectories a, b and c
will be discussed in Section 2.
}
\label{fig:phasediagram}
\end{figure}

Arguments based on a variety of 
models \cite{NJL,steph,ARW1,RappETC,bergesraj,stephetal,PisarskiRischke1OPT,CarterDiakonov}
indicate that the transition as a function of $T$ is first order at 
large $\mu$.  
This suggests that the
phase diagram features a critical point $E$ at which
the line of first order phase transitions present for 
$\mu>\mu_E$ ends, as shown in Figure 1.\footnote{If
the up and down quarks were massless, $E$ would
be a tricritical point, at which the first
order transition becomes second order.}
At $\mu_E$, the phase transition at $T=T_E$ is second order
and is in the Ising universality class \cite{bergesraj,stephetal}.
Although the
pions remain massive, the correlation length in the $\sigma$ channel
diverges due to universal long wavelength fluctuations
of the order parameter.
This results in characteristic signatures,
analogues of critical opalescence in the sense that they
are unique to collisions which freeze out near the
critical point, which
can be used to discover $E$ \cite{SRS1,SRS2}.   

The position of the critical point is, of course, not universal.
Furthermore, it is sensitive to the value of the strange
quark mass.  $\mu_E$ decreases as $m_s$ is decreased \cite{SRS1},
and at some $m_s^c$, it reaches $\mu_E=0$ and the transition
becomes entirely first order \cite{rajwil}.  
The value of $m_s^c$ is an open question,
but lattice simulations suggest that it is about half the
physical strange quark mass \cite{columbia,kanaya}, although 
these results are not yet conclusive \cite{oldkanaya}.
Of course, experimentalists cannot vary $m_s$.  They
can, however, vary $\mu$.  The AGS, with beam energy 11 AGeV
corresponding to
$\sqrt{s}=5$ GeV, creates fireballs which freeze out
near $\mu\sim 500-600$ MeV \cite{PBM}.  
When the SPS runs with $\sqrt{s}=17$ GeV
(beam energy 158 AGeV),
it creates fireballs which freeze out near $\mu\sim 200$ MeV \cite{PBM}.
RHIC will make even smaller values of $\mu$ accessible.
By dialing $\sqrt{s}$ and thus $\mu$, experimenters can find
the critical point $E$. 

\subsection{Detecting the Critical Point}

Predicting
$\mu_E$, and thus suggesting the $\sqrt{s}$ to
use to find $E$, is beyond the reach of present
theoretical methods
because
$\mu_E$ is both nonuniversal and sensitively dependent on the mass
of the strange quark.   
Crude models suggest that $\mu_E$ could be $\sim 600-800$ MeV
in the absence of the strange quark \cite{bergesraj,stephetal}; 
this in turn suggests that
in nature $\mu_E$ may have of order half this value, and may therefore
be accessible at the SPS if the SPS 
runs with $\sqrt{s}<17$ GeV.   However, at present theorists cannot
predict the value of $\mu_E$ even to within a factor of two.
The SPS can search a significant fraction of the parameter
space; if it does not find $E$, it will then be up to 
the RHIC experiments to map the $\mu_E< 200 $ MeV region.

Locating $E$ on the phase diagram can only be done convincingly by an
experimental discovery.  Theorists can, however, do reasonably well at
describing the phenomena that occur near $E$, thus enabling
experimenters to locate it.  This is the goal of Ref.  \cite{SRS2}.
The signatures proposed there are based on the fact that $E$ is a
genuine thermodynamic singularity at which susceptibilities diverge
and the order parameter fluctuates on long wavelengths. The resulting
signatures are {\it nonmonotonic} as a function of $\sqrt{s}$: as this
control parameter is varied, we should see the signatures strengthen
and then weaken again as the critical point is approached and then
passed.

The simplest observables to use are the event-by-event fluctuations of
the mean transverse momentum of the charged particles in an event,
$p_T$, and of the total charged multiplicity in an event, $N$.  One
analysis described in detail in Ref. \cite{SRS2} is based on the ratio
of the width of the true event-by-event distribution of the mean $p_T$
to the width of the distribution in a sample of mixed events. This
ratio was called $\sqrt{F}$. NA49 has measured $\sqrt{F}=1.002\pm
0.002$ \cite{NA49,SRS2}, which is consistent with expectations for
noncritical thermodynamic fluctuations.\footnote{In an infinite system
made of classical particles which is in thermal equilibrium,
$\sqrt{F}=1$.  Bose effects increase $\sqrt{F}$ by $1-2\%$
\cite{Mrow,SRS2}; an anticorrelation introduced by energy conservation
in a finite system --- when one mode fluctuates up it is more likely
for other modes to fluctuate down --- decreases $\sqrt{F}$ by $1-2\%$
\cite{SRS2}; two-track resolution also decreases $\sqrt{F}$ by $1-2\%$
\cite{NA49}. The contributions due to correlations introduced by
resonance decays and due to fluctuations in the flow velocity are each
significantly smaller than $1\%$ \cite{SRS2}.}  Critical fluctuations
of the $\sigma$ field, i.e. the characteristic long wavelength
fluctuations of the order parameter near $E$, influence pion momenta
via the (large) $\sigma\pi\pi$ coupling and increase $\sqrt{F}$
\cite{SRS2}.  The effect is proportional to $\xi_{\rm freezeout}^2$,
where $\xi_{\rm freezeout}$ is the $\sigma$-field correlation length
of the long-wavelength fluctuations at freezeout \cite{SRS2}.  If
$\xi_{\rm freezeout}\sim 6$ fm, the ratio $\sqrt{F}$ increases by
$10-20\%$, fifty times the statistical error in the present
measurement \cite{SRS2}.  This observable is valuable because data on
it has been analyzed and presented by NA49, and it can therefore be
used to learn that Pb+Pb collisions at 158 AGeV do {\it not} freeze
out near~$E$.

Once $E$ is located, however, other observables which 
are more sensitive to critical effects will be more useful.
For example, a $\sqrt{F_{\rm soft}}$,
defined using only the softest $10\%$ of the pions in each event, 
will be much more sensitive to the critical long wavelength 
fluctuations.  The higher $p_T$ pions are less affected
by the $\sigma$ fluctuations \cite{SRS2}, 
and these relatively unaffected pions
dominate the mean $p_T$ of all the pions in the
event.  This is why the increase in $\sqrt{F}$ near the critical point 
will be much less than that of $\sqrt{F_{\rm soft}}$. 

The multiplicity of soft pions is an 
example of an observable which may
be used to detect the critical fluctuations 
without an event-by-event analysis.
The post-freezeout decay of sigmas, which are copious
and light at freezeout near $E$ and which
decay subsequently when their mass increases above
twice the pion mass, should result in a population of pions 
with $p_T\sim m_\pi/2$ which appears only for freezeout
near the critical point \cite{SRS2}.  
If $\xi_{\rm freezeout}\gtrsim 1/m_\pi$, this population
of unusually low momentum pions will be comparable in
number to that of the ``direct'' pions (i.e. those which
were pions at freezeout) and will result in a large
signature. 

The variety of observables
which should all show nonmonotonic behavior near the critical
point is sufficiently great that if it were to turn out that 
$\mu_E<200$ MeV, making $E$ inaccessible to the SPS, all four
RHIC experiments could play a role in the study of the critical
point.

\subsection{How Large Can $\xi$ Grow?}

Our purpose in this paper is to estimate how large 
$\xi_{\rm freezeout}$ can become, thus making the predictions
of Ref. \cite{SRS2} for the magnitude of various signatures
more quantitative.  In an ideal system of infinite size
which was held at $T=T_E$; $\mu=\mu_E$ for an infinite time,
the correlation length $\xi$ would be infinite.  Ref. \cite{SRS2}
estimated that finite size effects limit $\xi$ to be 
about 6 fm at most. We will argue in this paper that
limitations imposed by the finite duration of a heavy
ion collision are more severe, preventing
$\xi$  from growing larger than about $2/T_E \sim 3$ fm.

\subsection{$T_E$, $T_{\rm freezeout}$, and $T_0$}

We will do the calculation in the next section in the three-dimensional
Ising model, as appropriate for describing the universal dynamics of
the long wavelength fluctuations near the critical point. However, in
order to relate a calculation in the Ising model to experiments which
explore the QCD phase diagram, we will need numerical values for three
temperature scales.  Several other nonuniversal quantities will also
enter our calculation; we will discuss them in the next section as
they arise.  We will see that in the end, only one combination of
nonuniversal quantities plays a role in our estimates.

We expect $T_E$ to be slightly less
than the temperature range at which the crossover
occurs at $\mu=0$.
We therefore take $T_E=140$ MeV, 
at the low end of lattice estimates for the 
$\mu=0$ crossover temperature.  

As we have discussed at length, we know very little about 
$\mu_E$. Fortunately, we will not need a numerical value 
for $\mu_E$ below.

Pb+Pb collisions at 158 AGeV freeze out at about 120 MeV,
and NA49 data \cite{NA49} demonstrate clearly that they do
{\it not} freeze out near $E$ \cite{SRS2}.  We also know \cite{SRS1}
that if the matter produced in a heavy ion collision
comes near $E$, the large specific heat characteristic
of $E$ will cause the system to ``linger'' --- the expansion
will cause the energy density to decrease as usual,
but this will result in an unusually slow temperature decrease.
The freezeout temperature is therefore expected to be unusually
close to the critical temperature for collisions which have the
appropriate $\mu$ to
pass near $E$.  For concreteness, we will take 
$T_{\rm freezeout}=130$ MeV.\footnote{Note that experimenters
do have some control over $T_{\rm freezeout}$. Using smaller
ions results in a fireball which freezes out earlier, at
a larger $T_{\rm freezeout}$ \cite{laterfreezeout}.}
(If the freeze-out temperature 
in Pb+Pb collisions at 158 AGeV is closer to 100 MeV,
as some authors estimate \cite{Heinz2}, then it may be 
better to estimate that collisions which pass near $E$
freezeout at $T_{\rm freezeout}=120$ MeV.)

Finally, we need to estimate $T_0$, the temperature
at which we can begin an Ising model treatment.
The three dimensional Ising model is only valid close
enough to $E$ that the correlation length $\xi > 1/T_E$.    
In this critical region,
the long wavelength fluctuations of the order parameter
become effectively three dimensional.  (We will find
that $\xi$ is never $\gg 1/T_E$.  This means that
our estimates are not precise.)
We need to know how far above $T_E$ the equilibrium
correlation length is larger than $1/T_E$.  
The model of Ref. \cite{bergesraj}
suggests that $\xi_{\rm eq} > 1/T_E$ for $(T-T_E)/T_E \lesssim 0.2 - 0.4$.
This estimate is based on a mean field analysis of
a toy model, and so should not be taken too seriously.
For concreteness we shall assume that 
$\xi_{\rm eq} = 1/T_E \equiv \xi_0$ 
at $T_0 = 180$ MeV, 40 MeV above $T_E\sim 140$.
We will use $\xi_0=1.4$ fm to 
set the scale below, in the sense that we will estimate
the factor by which $\xi/\xi_0$ grows as the system cools.
$\xi_0=1.4$ fm is simply a definition;
$T_0$, the temperature at which the equilibrium correlation
length $\xi_{\rm eq}=\xi_0$, is a quantity which must be estimated
and which will affect our results.

\section{Slowing Out of Equilibrium}

The nonequilibrium dynamics which we analyze in this
paper is fundamental in the sense that it is {\it guaranteed}
to occur in a heavy ion collision which passes near $E$,
even if local thermal equilibrium is achieved 
at a higher temperature during the earlier evolution
of the plasma created in the collision.  
We assume early thermal (although not necessarily chemical)
equilibration, and ask how the system evolves
out of equilibrium as it passes $E$.  More precisely,
we will assume that when the system has cooled
to $T=T_0=180$ MeV, it is in equilibrium, with
$\xi(T_0)=\xi_{\rm eq}(T_0)=\xi_0$.
For the present, assume that the system cools 
through the critical point $E$, as sketched in trajectory (a)
of Figure 1.
If it were to cool arbitrarily slowly, $\xi=\xi_{\rm eq}$
would be maintained at all temperatures, and $\xi$ would
diverge at $T_E$.  However, it would take an infinite
time for $\xi$ to grow infinitely large.  Indeed, near
a critical point, the long correlation length results
in long equilibration times, a phenomenon known 
as critical slowing down.   This means that the
correlation length cannot grow as fast as $\xi_{\rm eq}$,
and the system cannot stay in equilibrium.  

We describe the effects of critical slowing down
on the time development of the correlation length $\xi(t)$ 
using the following equation for $m_\sigma(t)\equiv 1/\xi(t)$:
\begin{equation}
\frac{d}{dt} \, m_\sigma(t) = - \Gamma\left(m_\sigma(t)\right)
\left(m_\sigma(t)-\frac{1}{\xi_{\rm eq}(t)}
\right)\ .
\label{mainequation}
\end{equation}
Here, $\Gamma$ parametrizes the rate at which an out-of-equilibrium
value of $m_\sigma$ approaches its equilibrium value.  If 
$m_\sigma$ is close to its equilibrium value, the theory
of dynamical critical phenomena \cite{HoHa} tells us
that 
\begin{equation}
\Gamma(m_\sigma) = \frac{A}{\xi_0} \, \left(m_\sigma \xi_0\right)^z
\label{Gammadef}
\end{equation}
where $z$ is a universal exponent and we have used
$\xi_0$ to set the scale, making $A$ a dimensionless 
constant.  Knowing that we are interested in a system
which is in the same static universality class as the
3-dimensional Ising model is {\it not} enough to tell
us $z$. There are in general several different
dynamical universality classes corresponding to
a given static universality class.
However, knowing in addition that: (i) the chiral order
parameter is not a conserved quantity; (ii) there
are other conserved quantities in the system, like
the baryon number density; and (iii)
there are no Poisson bracket relations between the
order parameter and the conserved quantities, tells
us that our system belongs in the dynamical universality
class named Model C in 
Halperin and Hohenberg's classification \cite{HoHa}
of dynamical critical phenomena, 
and has 
\begin{equation}
z=2+\alpha/\nu\approx 2.17\ ,
\end{equation}
where we have taken $\alpha=0.11$ and $\nu=0.630$ from
Ref. \cite{GuidaZinn}.
The dimensionless constant $A$ is nonuniversal.
We have no way to estimate it other than to guess
that it is of order 1.
We will explore the sensitivity of our results
to different choices of $A$ below.

We will use the differential equation (\ref{mainequation})
to analyze how critical slowing down prevents
the correlation length $\xi$ from ``tracking''
$\xi_{\rm eq}(T(t))$.  Critical slowing down
guarantees that  the system falls out of equilibrium.
Note that the differential equation has only
been derived for small departures from equilibrium;
once $m_\sigma - \xi_{\rm eq}^{-1}$ is not small,
its use is not quantitatively 
justified.\footnote{For example, one might 
try the equation $d\xi/dt = - \Gamma (\xi -\xi_{\rm eq})$, instead of 
the equation (\ref{mainequation})
for $dm_\sigma/dt$.  These two equations give the same results 
for small departures from equilibrium, but they do not
agree in all circumstances. For example, in a system which
is not cooling and which has $T=T_E$ and $\xi_{\rm eq}=\infty$
for all time, only (\ref{mainequation}) yields the
correct result, namely $m_\sigma(t)\sim t^{-1/z}$ at late
time.}

We have initial conditions for the differential
equation (\ref{mainequation}), namely $m_\sigma(0)=1/\xi_0$.
Therefore, all we need in order to solve it
is a description of $\xi_{\rm eq}(t)$. This
requires $\xi_{\rm eq}(T)$, which we discuss
below, and also requires a description
of the cooling $T(t)$.  
This can be estimated
using hydrodynamic and cascade model calculations,
although these describe $T(t)$ assuming the
plasma is {\it not} cooling near the critical point $E$.
Hydrodynamic models (see, e.g., 
Refs. \cite{HS,Heinz})
describe $T(t)$ at central rapidity in the center
of mass frame via
\begin{equation}
\frac{dT}{dt} = - \frac{1}{\kappa} \frac{T}{t_0}\ .
\label{dTdt}
\end{equation}
Since we are only interested in a relatively small
range of temperatures around $T_E$, it will suffice
for us to treat $dT/dt$ as constant in time.  We discuss
the effects of the time dependence of $dT/dt$ below.
The expression (\ref{dTdt}) assumes that the $T$-dependence of
the energy density is $\epsilon \sim T^\kappa$ as in
a resonance gas, for which $\kappa\approx 6$ \cite{OldShuryak}.
The timescale $t_0$ is not constant over the whole history
of the collision. 
A simplified  estimate (made by
equating $t_0$ with the scattering time) suggests 
that in Pb+Pb collisions at 158 AGeV, $t_0$ is
between 4 and 10 fm at times of interest to us \cite{Heinz}.
This suggests  $-dT/dt=(2-6) {\rm ~MeV}/{\rm fm}$ at $T=140$ MeV.  
Careful analysis favors $t_0$ closer to $4$ fm \cite{Heinz2}. 
This agrees with a recent analysis
of these collisions using the URQMD cascade model, which
suggests $-dT/dt\approx 5{\rm ~MeV}/{\rm fm}$ at $T=140$ MeV \cite{Bravina}.
These estimates are all for cooling through $T=140$ MeV at
a $\mu$ such that one is not near the critical point.
As we discussed above, the cooling rate 
is likely to be unusually low near $E$
because of the large specific heat there; we will 
therefore take $-dT/dt \sim 4 {\rm ~MeV}/{\rm fm}$
as our estimate, noting also that the cooling rate
at RHIC will be slower still.

We wish to
use the three-dimensional Ising model to describe 
$\xi_{\rm eq}(T,\mu)$ near $E$.
In the Ising model, the order parameter $M$ (the magnetization)  
and the correlation
length $\xi$ are functions of the reduced temperature 
$r$ and the magnetic field $h$.  (In the Ising model,
$r$ is defined as $(T-T_c)/T_c$ and is
usually called $t$;
we reserve the symbol $t$ for time, however.)   The critical
point is at $r=h=0$; at this point, $M=0$ and $\xi_{\rm eq}=\infty$.
For $r<0$, there is a first order phase transition 
as a function of $h$ at $h=0$ between $M=|r|^\beta$ for
$h=0+$ and  $M=-|r|^\beta$ for $h=0-$. The exponent 
is $\beta=0.326$ for the 3-dimensional Ising model \cite{GuidaZinn}.
For $r>0$, $M$ increases
smoothly through zero as $h$ goes from negative to positive.
For $r=0$, the order parameter 
is $M= {\rm sign}(h) |h|^{1/\delta}$, with $\delta = 4.80$ \cite{GuidaZinn}.

We can now discuss how the Ising model $r$- and $h$-axes are mapped
onto the $(T,\mu)$ plane.  The $r$-axis is the direction tangential to
the line of first order phase transitions ending at $E$.  This is
shown in Figure 1.  There is no guarantee that the $h$-axis is
perpendicular to the $r$-axis when both are mapped onto the $(T,\mu)$
plane.  This mapping will in general deform the Ising axes, but we
have no way of estimating this deformation.\footnote{In the
electroweak phase diagram as a function of $T$ and Higgs mass $m_H$,
there is also a line of first order phase transitions ending at an
Ising critical point.  Here, the explicit mapping between Ising axes
and the $(T,m_H)$ plane has been constructed \cite{Tsypin}.  This is
possible only because there are reliable numerical methods for
analyzing the full, nonuniversal theory in the $(T,m_H)$ plane.
Universality arguments alone, which is all that we have at our
disposal in the absence of lattice simulations at nonzero $\mu$, do
not tell us how the Ising axes should be deformed in the $(T,\mu)$
plane of Figure 1.}  For simplicity, we draw the $h$-axis
perpendicular to the $r$-axis in Figure 1.  In thinking through the
mapping between QCD and the Ising model even qualitatively, it is
important to note that the QCD order parameter (
the chiral condensate $\langle \bar q q \rangle$) 
is offset with respect to the Ising model order parameter
(the magnetization $M$).  In the Ising model, $M=0$ at the critical
point and along the first order line one has phase coexistence between
phases whose $M$'s are equal in magnitude and opposite in sign.  In
QCD, $\langle \bar q q \rangle \neq 0$ at the critical point $E$,
because of the explicit breaking of the $O(4)$ symmetry by quark mass
terms.  
Near $E$, $\langle \bar q q \rangle$ corresponds to $M$ plus
an offset, and the phase coexistence is between phases with differing
values of $\langle \bar q q \rangle$ which both have the same sign.
In Figure 1, we take the $-h$ side of the Ising coexistence line to
correspond to the higher temperature side of the QCD coexistence line,
so that increasing $M$ corresponds to increasing the magnitude of
$\langle \bar q q \rangle$.

The matter created in heavy ion collisions at SPS energies will
follow a trajectory in the $(T,\mu)$ plane which is approximately
vertical as it cools. (See, for example, Ref. \cite{Bravina}.)  
We therefore begin by considering trajectory (a)
of Figure 1, which follows the 
$h$-axis, as this is likely not a bad approximation
to cooling at almost constant $\mu$.\footnote{Note that the $r$-direction,
corresponding to the reduced temperature direction in the Ising model,
is almost perpendicular to the $T$ direction in QCD. This is 
another reason why we have labeled it by a letter other than $t$.}
We have analyzed trajectories which pass through
$E$ at a variety of angles, for example like trajectory (b) in Figure~1.
The results do not differ qualitatively from those we present
in detail for a trajectory along the $h$-axis, unless   
the trajectory  passes through $E$ almost parallel to the $r$-axis.
At the end of
this section, we will present results for trajectories like (c) in 
Figure 1, which miss $E$ but
come close to it.

Let us take the initial temperature in our calculation,
$T=T_0=180$ MeV, to correspond to $h=h_0=-0.2$. 
Along the $h$-axis,\footnote{Along the $r$-axis, $\xi_{\rm eq}\sim r^{-\nu}$;
along the $h$-axis,   $\xi_{\rm eq}\sim h^{-\nu/\beta\delta}$;
for trajectories like (b) of Figure 1 which pass through $E$
at generic angles, the larger exponent $(\nu/\beta\delta)$
is the relevant one.  Our $h$-axis analysis is therefore
a good guide to the generic case.}
as in trajectory (a),
the equilibrium correlation length
is a power law in $h$:
\begin{equation}
\xi_{\rm eq}(h) =  \left|\frac{h}{h_0}\right|^{-\nu/\beta\delta}\ ,
\end{equation}
where we have normalized $\xi$ by
setting $\xi_{\rm eq}(h_0)=1$.  That is,
we measure $\xi$ in units of $\xi_0=1.4$ fm.
With units chosen, we can now rewrite the equation (\ref{mainequation})
which describes the dynamics of the growth of the correlation
length $\xi=1/m_\sigma$ in terms of Ising model variables as
\begin{equation}
\frac{d}{dh} \, m_\sigma(h) =  - a\, \Biggl(m_\sigma(h)\Biggr)^z \,
\left(m_\sigma(h) - 
\frac{1}{\xi_{\rm eq}(h)}\right)
\label{universalequation}
\end{equation}
where the nonuniversal constant $a$ is
related to the other nonuniversal parameters we have discussed by:
\begin{equation}
a= A \left(\frac{dh}{dt}\right)^{-1} = 
A \left(\frac{h_0}{T_0-T_E}\frac{dT}{dt}\right)^{-1}\ .
\end{equation}
Nonuniversal parameters 
appear in equation (\ref{universalequation})
only in the single combination $a$.
Taking the nonuniversal constant from (\ref{Gammadef}) to be 
$A\sim 1$, using $(T_0-T_E)=40$ MeV, 
$dT/dt=-4 {\rm ~MeV}/{\rm fm}$ and $h_0=-0.2$ yields
the estimate
\begin{equation}
a \sim 50 \ .
\end{equation}
In fact, because $\xi_{\rm eq}(h)$ is a power law in $h$,
if one changes $h_0$ and then redefines 
the units of $\xi$ so that $\xi_{\rm eq}(h_0)$ is again set to one, 
equation (\ref{universalequation}) is unaffected.  Our results
are therefore determined solely by $A$,  $(T_0-T_E)$ and $dT/dt$
in the single combination $a$, together with the assumption that
the system begins in equilibrium at $T=T_0$.

\begin{figure}[t]
\begin{center}
\epsfig{file=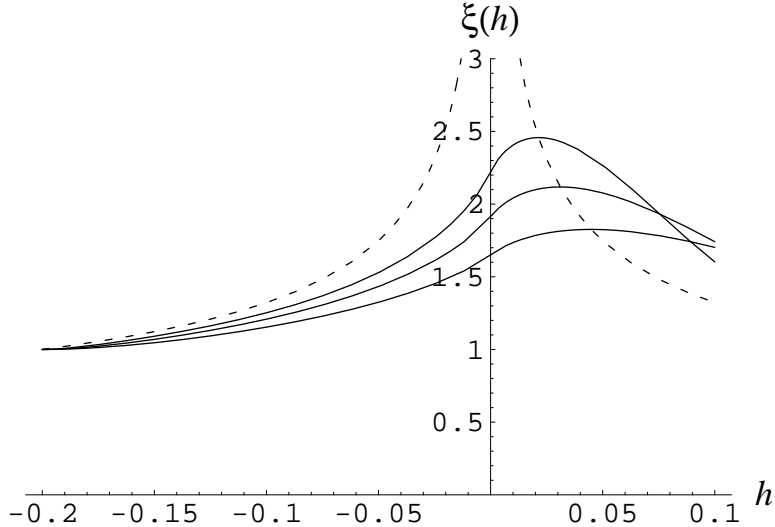,width=4.5in}
\end{center}
\vspace{-0.2in}
\caption{Behavior of the correlation length for 
cooling through the critical point along the $h$-axis
of Figure 1.  The equilibrium correlation length
is shown as a dashed line.  The true correlation
length is shown for (bottom to top) $a=25,50,100$.
Our units, described in the text, are such that $h=-0.2,-0.1,0,0.1$
corresponds to $T=180,160,140,120$ MeV, and $\xi$ is measured
in units of $\xi_0 = 1.4$ fm.  
}
\label{fig:haxis}
\end{figure}
We can now use (\ref{universalequation})
to learn how much $\xi$ grows relative to $\xi_0 = 1.4$ fm.
Given the uncertainties in the
determination of $a$, in Figure 2 we show $\xi(h)$ 
obtained by solving (\ref{universalequation}) for $a=25,50,100$.
Four lessons are apparent:

First, critical slowing down has a large effect.  Although
by assumption we begin in thermal equilibrium with $\xi=\xi_{\rm eq}$
at $T=T_0$, the fact that the dynamics slows down in the vicinity
of $E$ prevents $\xi$ from tracking $\xi_{\rm eq}$ and
growing very large.  

Second, our
results do not depend sensitively on the parameter $a$.  This
is fortunate, since there are so many uncertainties involved
in estimating $a$.  For $a=25,50,100$, the maximum correlation
length which is achieved is $1.8\xi_0$, $2.1\xi_0$, $2.5\xi_0$,
corresponding to 2.6, 3.0, 3.4 fm.  This means that although
our estimate is only qualitative, it is clear that $\xi$
cannot grow as large as 6 fm.  (To obtain a maximum value of 
$\xi=4 \xi_0$ would require $a=1000$.  Although $a$ is uncertain,
this large a value seems out of the question.) 
We estimate that $\xi$ grows to about twice $\xi_0$,
corresponding to approximately 3 fm.  

Third, since previous work \cite{SRS2}
suggests that finite size effects limit $\xi$ to $\xi<6$ fm, we
conclude that slowing out of equilibrium (i.e. the combination
of finite time and critical slowing down) imposes the more
stringent constraint on $\xi$.  We have analyzed this 
nonequilibrium effect as if the system were spatially homogeneous.
Had we found correlation lengths growing beyond 6~fm,
we would have to do a much more complicated analysis, taking
both the finite time and the inhomogeneous spatial dynamics
into account.  Since we find that $\xi$ only grows to about 3 fm,
this is not necessary.

Fourth, just as critical slowing down prevents $\xi$ from growing
as fast as $\xi_{\rm eq}$ does, it also prevents $\xi$ from
shrinking as fast as $\xi_{\rm eq}$ does after $E$ has 
been passed.  Whether we estimate $T_{\rm freezeout}\sim 130$ MeV
(corresponding to $h=.05$) or $\sim 120$ MeV ($h=0.1$)
for trajectories passing near $E$, 
one finds $\xi_{\rm freezeout}\sim 2 \xi_0$.
We can also argue that even if an $a$ as large as 1000 were
possible, our conclusions would be little affected:
If $a=1000$, and 
$\xi$ follows $\xi_{\rm eq}$ closely enough that it 
increases to $4\xi_0$, $\xi$ also tracks $\xi_{\rm eq}$
more closely as it decreases
below $T_E$.  
If $a=1000$, it turns out that 
$\xi$ is quite close to $\xi_{\rm eq}$ by the time
$h=.05$.
Thus, although increasing $a$ to a ridiculous
extent does increase the maximum value of $\xi$, it 
has little effect on $\xi_{\rm freezeout}$.
This is further evidence that although
our estimate $\xi_{\rm freezeout}\sim 2\xi_0 \sim 3$ fm
is qualitative, it is robust.

The $a$-dependence of the maximum value attained by $\xi$
can be understood analytically at large $a$.  
For large $a$, $m_\sigma$ tracks its equilibrium value
$m_\sigma^{\rm eq}=|h/h_0|^{\nu/\beta\delta}$  
well until $|h/h_0|$ is quite small.  If we define 
$\epsilon \equiv m_\sigma - m_\sigma^{\rm eq}$, we
can use (\ref{universalequation}) to show that
$\epsilon < m_\sigma^{\rm eq}$ as long as 
\begin{equation}
\frac{h}{h_0} > \left( \frac{\nu}{\beta\delta} \frac{1}{a} 
\right)^{\frac{1}{1+z\nu/\beta\delta}}\ .
\end{equation}
If we assume that once $\xi$ begins to drop out of 
equilibrium (i.e. once $\epsilon$ begins to grow 
comparable to $m_\sigma^{\rm eq}$), little further
growth of $\xi$ occurs before $\xi$ reaches its maximum,
we predict that $\xi$ will peak at
\begin{equation}
\xi^{\rm max} = \left(c\,a\right)^{\frac{\nu/\beta\delta}
{1+z\nu/\beta\delta}} = \left(c\,a\right)^{0.215}\ ,
\label{scaling}
\end{equation}
for some constant $c$.  The maxima of solutions 
to (\ref{universalequation}) obtained numerically 
follow this scaling relation (with $c=0.65$) quite
accurately once $a>1000$ or so.
Even at much smaller $a$, as in Figure 2, $\xi^{\rm max}$ is
within a few percent of that in (\ref{scaling}).
This scaling relation 
explains why our results are so weakly dependent on $a$.
Note that even with the scaling relation in hand, 
full solutions as in Figure 2 are of value because they
allow us to estimate $\xi_{\rm freezeout}$ and not just $\xi_{\rm max}$.

We have to this point assumed that $dT/dt$ is approximately
constant as the system cools through $T_E$.  This is
an oversimplification. It is more reasonable
to assume that $ds/dt$ is approximately constant, where
$s$ is the entropy density.  Since
\begin{equation}
\frac{dT}{dt} = \frac{1}{C_V}\frac{ds}{dt}\ 
\end{equation}
and the specific heat $C_V$ is peaked at $T_E$, we
expect that $dT/dt$ is unusually small
near $T_E$. As we discussed
above, this ``lingering'' results in a $T_{\rm freezeout}$
which is unusually close to $T_E$ \cite{SRS1}.
Here, we estimate the effect of lingering near $E$ on
the growth of $\xi$.
Along the $h$-axis, the specific heat due
to the long wavelength sigma fluctuations
diverges like $C_V \sim h^{-\gamma/\beta\delta} 
\sim \xi_{\rm eq}^{\gamma/\nu}$ 
in thermal equilibrium \cite{SRS1}.  
The exponent $\gamma=1.240$ \cite{GuidaZinn}.
We take 
$C_V = c_1 + c_2\xi^{\gamma/\nu}$, where $c_1$ is the specific heat
due to all the degrees of freedom other than the sigma and is
smooth near $T_E$.  Note that
$C_V$ depends on the actual correlation length
$\xi$, and not on $\xi_{\rm eq}$. In our
dynamical nonequilibrium setting, therefore, $C_V$ 
peaks but does not diverge.
We can implement lingering
in our calculation by replacing the constant $a$ 
in (\ref{universalequation}) by
\begin{equation}
a(h) = a\left[ (1-b) + b\left[ m_\sigma(h)\right]^{-\gamma/\nu}\right]\ .
\label{bdef}
\end{equation}
Here, $a$ is the same constant as before and 
the constant $b=c_2/(c_1+c_2)$ 
is the fraction of the specific heat at $T=T_0$ which
is due to sigma fluctuations.
This fraction is 
perhaps about $0.1$ and is surely less than $0.25$.
As the system cools
from $T_0$ to $T_E$, the sigma contribution to $C_V$ 
grows and peaks.   We find that 
changing constant $a$ to $a(h)$ as in (\ref{bdef}) 
with $b=0.25$ increases $\xi_{\rm max}$ 
by about $10\%$ beyond that shown in Figure 2, 
and increases $\xi_{\rm freezeout}$
by somewhat less.  For $b=0.1$, the increase in $\xi_{\rm max}$
is about $5\%$.
We conclude that because $C_V$ receives contributions
from all degrees of freedom and not just from the sigma fluctuations,
and because $C_V$, like $\xi$, peaks but does not diverge, 
the reduction in $dT/dt$ near $E$ is not large enough
to significantly increase $\xi$ beyond our previous estimates.

We now ask how much our results change if we consider trajectories
like (c) in Figure 1
which come close to, but miss, $E$.  Our analysis
can easily be extended to cover those trajectories
which pass $E$ on the crossover side ($r>0$; $T<T_E$).
In an appendix, we present the Ising model 
expression for $\xi_{\rm eq}(r,h)$ near the critical point.
We use this expression to evaluate $\xi(h)$ for trajectories parallel
to the $h$-axis with $r=0.12$, $r=0.19$ and $r=0.33$, for which 
$\xi_{\rm eq}$ peaks at $4\xi_0$, $3\xi_0$ and $2\xi_0$.   
The results are shown in Figure 3  in which we have 
taken $a=50$.  
Note that in plotting Figure 3 
we have defined $\xi_0=\xi(r,h_0)=1$
anew for each $r$.  
We see that as long as $\xi_{\rm eq}$ peaks
at $3\xi_0$ or higher, the dynamics of $\xi$ is almost
the same as for the trajectory of Figure 2 which goes
precisely through $E$. Even for a trajectory which
misses $E$ by enough that $\xi_{\rm eq}$ peaks at 
only $2\xi_0$, the actual correlation length $\xi$
grows by a factor which is within $20\%$ of that for
trajectories which pass arbitrarily close to $E$.
Just as the growth of $\xi$ is robust with respect 
to changes in $a$, it is robust with respect to how
close the trajectory comes to $E$, for those trajectories
which come close enough.
\begin{figure}[t]
\begin{center}
\epsfig{file=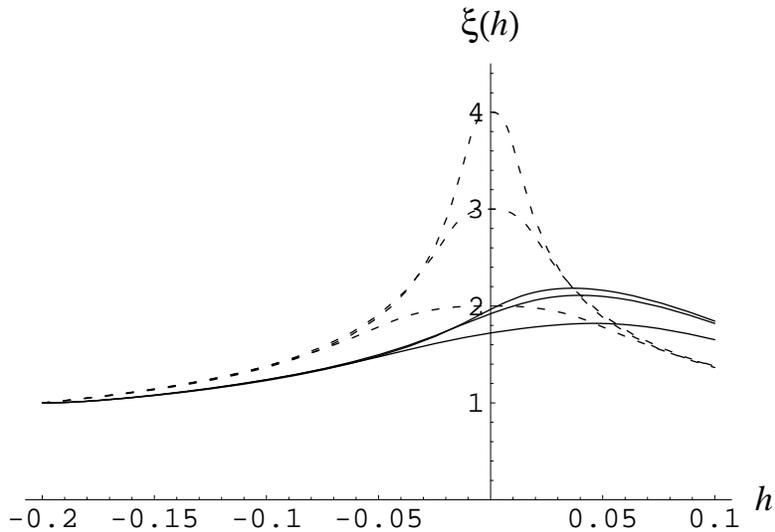,width=4.5in}
\end{center}
\vspace{-0.2in}
\caption{The dashed curves show $\xi_{\rm eq}(h)$ for
trajectories with (top to bottom)
$r=0.12,\ 0.19,\ 0.33$.  The solid curves show
the corresponding nonequilibrium correlation lengths $\xi$,
assuming $a=50$. Curves with different $r$ have each been
normalized to begin at $\xi(h_0)=1$.
} 
\label{fig:nonzeror}
\end{figure}

Our analysis 
is not sufficient to describe the dynamics
for those trajectories which pass to the first order side of $E$,
because we do not treat the dynamics of bubble nucleation and 
phase coexistence.
Near enough to $E$, though, the first order transition is so
weak that it will not have detectable effects given the finite
length and time scales in a heavy ion collision, and the physics
is likely qualitatively similar to that we have analyzed
on the crossover side of $E$. 
Farther from $E$ on the first order side, 
this is not the case.  Farther from $E$, though,
the correlation length is never large.

\section{Consequences}

Our results have a number of consequences which should
be taken into account both in planning experimental
searches for the QCD critical point, and in planning future
theoretical work.

Because of critical
slowing down, the correlation length in a heavy ion
collision cannot grow
as fast as it would in equilibrium; this means that
$\xi_{\rm freezeout}$ is likely about 3 fm for trajectories
passing near $E$. Although finite size effects 
alone would allow a correlation length as large as 6 fm,
this is unrealistic to expect in a heavy ion collision.
This effect arises due to {\it guaranteed} nonequilibrium
physics:  even if heavy ion collisions achieve local
thermal equilibrium above the transition, as we have
assumed, if they cool through the transition near
the critical point they must ``slow out of equilibrium.''
By this we mean that the correlation length cannot grow
as it would in equilibrium, because the long
wavelength dynamics are slow near~$E$.

Critical slowing down
also prevents the correlation length from decreasing quickly after
passing the critical point.  One therefore need not worry about $\xi$
decreasing significantly between the phase transition and 
freezeout.

One need not hit $E$ precisely in order
to find it.  The results shown in Figure 3 demonstrate that if one were to
do a scan with collisions at many finely spaced values of
the energy and thus $\mu$, one would see signatures of $E$
with approximately the same magnitude over a broad range
of $\mu$.  The magnitude of the 
signatures will not be narrowly peaked as $\mu$ is varied.
As long as one gets close enough to $E$ that the equilibrium
correlation length is $(2-3)\xi_0$, the actual correlation
length $\xi$ will grow to $\sim 2\xi_0$.  There is no advantage
to getting closer to $E$, because critical slowing down
prevents $\xi$ from getting much larger even if $\xi_{\rm eq}$ does.
Data at many finely spaced values of $\mu$ is
{\it not} called 
for.\footnote{Analysis within the toy model of Ref. \cite{bergesraj}
suggests that in the absence of the strange quark, the
range of $\mu$ over which $\xi_{\rm eq}> 2$ fm is about
$\Delta \mu \sim 120$ MeV for $\mu_E \sim 800$ MeV.
Similar results can be obtained \cite{Misha} within 
a random matrix model \cite{stephetal}.  It is likely
over-optimistic to estimate $\Delta \mu \sim 120$ MeV
when the effects of the strange quark are included
and $\mu_E$ itself is reduced. A conservative estimate would
be to use the models to estimate that $\Delta \mu/\mu_E \sim 15\%$.}

Only one combination of the nonuniversal 
quantities (called $a$ above) plays an important role in estimating the
dynamics of $\xi$. The uncertainty in $a$ is the sum of
that in its three factors: $A$ (the
nonuniversal constant in the dynamical scaling law (\ref{Gammadef})),
$dT/dt$ and $T_0-T_E$. It is already fortunate that only
one combination $a$ matters; it is even more fortunate that
our results are not very
sensitive to the value of $a$.  This means that
although our results are not completely quantitative,
they are robust.  In addition to the uncertainty in $a$, however,
our results cannot be treated as precise because the 
QCD dynamics are precisely described by the three-dimensional
Ising model dynamics only if $\xi\gg 1/T_E$, and we have
found that $\xi$ does not grow beyond $\sim 2/T_E$.

There are a number of steps
that could be taken in future work to refine
our estimate.  One could do a more complete job
of analyzing the universal dynamics of a system
which passes near an Ising critical point.
For example, instead of simply writing a differential
equation for $\xi$, one could follow the full 3+1-dimensional
dynamics in a Langevin simulation, from which one
would measure $\xi$.  Doing this, however, would still
leave one facing the same nonuniversal
uncertainties 
which we face in our treatment. If we simply ask how
to reduce the uncertainty in $a$, perhaps the hardest
part of this task would be a 
reliable calculation of $A$, as that would
require a reliable calculational method for QCD dynamics
at nonzero $T$ and $\mu$.   
The other two ingredients
in $a$ are 
likely to become better known as the modeling
of heavy ion collisions and the analysis of data from
these collisions proceed.  It seems, though, that the uncertainty
in $A$ will prevent a fully quantitative calculation 
of $a$ for the foreseeable future.
Our results are sufficiently insensitive
to $a$ that they suffice to estimate the magnitude
of signatures; when these signatures are found, perhaps
they will give us more quantitative information
about the nonuniversal quantities which go into $a$.

Knowing that we are
looking for  $\xi_{\rm freezeout} \approx 3$ fm is very helpful
in suggesting 
how to employ the signatures described in detail
in Ref. \cite{SRS2}.    The excess of pions with $p_T\sim m_\pi/2$
arising from post-freezeout decay of sigmas is large 
as long as $\xi_{\rm freezeout}\sim 1/m_\pi$, and does not
increase much further if $\xi_{\rm freezeout}$ is longer.
This makes it an ideal signature.
The increase in the
event-by-event fluctuations in the mean transverse
momentum of the charged pions in an event (described
by the ratio $\sqrt {F}$ of Ref. \cite{SRS2}) is proportional
to $\xi_{\rm freezeout}^2$.  The results of Ref. \cite{SRS2} suggest that
for $\xi_{\rm freezeout}\sim 3$ fm, this will be a $3-5\%$ effect.
This is ten to twenty times larger than the statistical error in the
present NA49 data, but not so large as to make one confident
of using this alone as a signature for $E$.   The solution
is to use signatures which focus on the event-by-event
fluctuations of only the low momentum pions. Unusual
event-by-event
fluctuations in the pion momenta arise via the coupling
between the pions and the sigma order parameter which, at freezeout, is
fluctuating with correlation length $\xi_{\rm freezeout}$.
This interaction has the largest effect on the softest 
pions \cite{SRS2}.
$\sqrt{F_{\rm soft}}$,
described in the introduction, is a good example of an observable
which takes advantage of this.  Depending on the details of the
cuts used to define it, it should be enhanced by many tens
of percent in collisions passing near $E$.
Ref. \cite{SRS2} suggests other such observables,
and more can surely be found.  Together, the excess
multiplicity at low momentum (due to post-freezeout sigma decays) and
the excess event-by-event fluctuation of the momenta of
the low momentum pions (due to their coupling to the order parameter
which is fluctuating with correlation length $\xi_{\rm freezeout}$)
should allow a convincing detection of the critical point $E$.
Both should behave nonmonotonically as the collision energy,
and hence $\mu$, are varied.  Both should peak for those
heavy ion collisions which freeze out near $E$, with
$\xi_{\rm freezeout}\sim 3$ fm.

\vspace{3ex}
{\samepage 
\begin{center} Acknowledgments \end{center}
\nopagebreak

We acknowledge helpful conversations with J. Bowers,
B. Halperin, U. Heinz, E. Shuryak and M. Stephanov.
This work is supported in part  by the U.S. Department
of Energy (D.O.E.) under cooperative research agreement \#DF-FC02-94ER40818.
The work of KR is supported in part by a DOE OJI Award and by the
Alfred P. Sloan Foundation.
}

\appendix

\section{Equilibrium Correlation Length}

In this appendix we present the equations that we used to calculate
the equilibrium correlation length $\xi_{\rm eq}$ in the critical
region as a function of reduced temperature $r$ and external magnetic
field $h$.  We use Widom's scaling form \cite{Brez}
\begin{equation}
\xi_{\rm eq}^2 (r, M) = f^2 
\,M^{-2 \nu / \beta} g \left({|r| \over |M|^{1/\beta}}
\right ) ,
\label{xi-scaling}
\end{equation}
in which $M$ is the magnetization and $\nu = 0.630$ and $\beta = 0.326$
are the three-dimensional 
Ising model critical exponents \cite{GuidaZinn}. The
$\ep$-expansion of $g(x)$ is given in \cite{Brez} to order $\ep^2$:
\begin{eqnarray}
g(x) = g_\ep(x) &=& 6^{-2 \nu} z \Biggl\{ 1 - {\ep \over 36} \left[ (5 +
6 \ln 3 ) z - 6 (1+z) \ln z \right] + \\
&& \ep^2 \Biggl[ {1 + 2 z^2 \over 72} \ln^2 z + \Bigl( {z \over 18} (z -
{1 \over 2} )(1 - \ln 3) - \nn
&& {1 \over 216} (16 z^2 - {47 \over 3} z - {56 \over 3}) \Bigr) \ln z +
\nn
&& {1 \over 216} \Bigl({101 \over 6} + {2 \over 3} I + 6 \ln^2 3 + 4 \ln 3
- 10 \Bigr) z^2 - \nn
&& {1 \over 216} \Bigl(6 \ln^2 3 + {44 \over 3} \ln 3 
+ {137 \over 9} + {8 \over 3} I \Bigr) z \Biggr] \Biggr\}, \nn
\mbox{where } z &\equiv & {2 \over 1 + {x \over 3}}, \mbox{ and } I = \int_0^1
{\ln \left[x(1-x)\right] \over 1 - x (1-x)} dx \approx -2.344 \nonumber.
\label{g-epsilon}
\end{eqnarray}
$f$ in (\ref{xi-scaling}) is a non-universal normalization
constant, often set to one.  Our choice of $f$ and
thus of units for $\xi$
is described in Section 2. 
The expression (A.2) is valid everywhere on the ($r,M$)
plane except the region of large $x$ (or, equivalently, $r\gg |M|^{1/\beta}$).
The correct result at large $x$ is
\begin{equation}
g_{\rm large}(x) = \left ({1 \over 3+x} \right )^{2 \nu},
\label{glarge}
\end{equation}
and we therefore construct a function $g(x)$ which smoothly interpolates
between $g_{\rm large}(x)$ at large $x$ and $g_\ep(x)$ at 
smaller $x$.  The only remaining difficulty is at $r\geq 0$, $M=0$.
Although the scaling form (\ref{xi-scaling}) 
with (\ref{glarge}) is well-behaved
in the $M\rightarrow 0$ limit, and yields 
\[
\xi_{\rm eq}(r \ge 0, M\rightarrow 0) = f |r|^{-\nu} ,
\]
at $M=0$ the scaling form is indeterminate
and one must impose the condition
$\xi_{\rm eq}(r \ge 0, M=0)=\xi_{\rm eq}(r \ge 0, M\rightarrow 0)$.


\begin{figure}[t]
\vspace{-0.4in}
\begin{center}
\epsfig{file=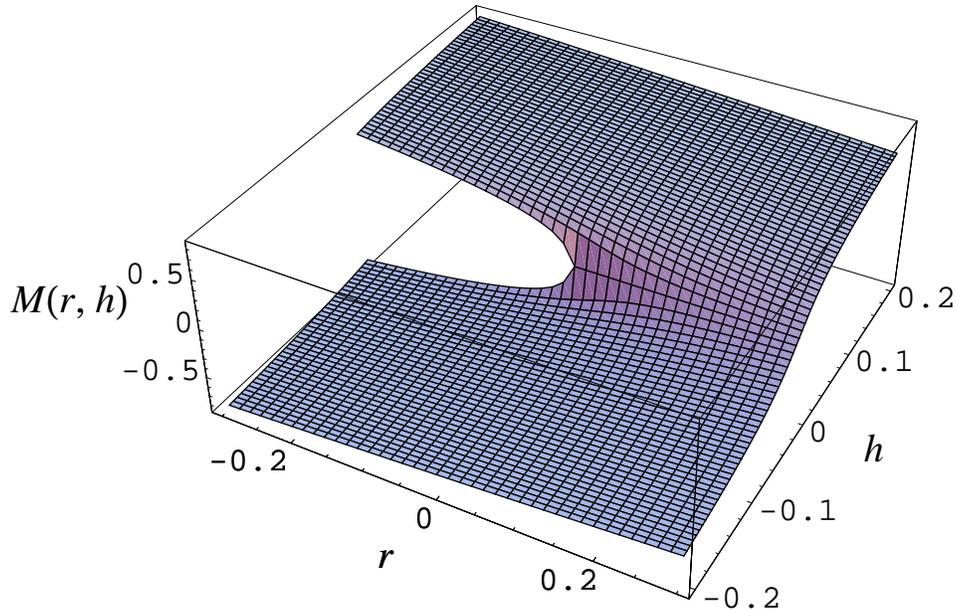,width=5.in}
\end{center}
\vspace{-0.5in}
\caption{Order parameter (magnetization) in the 3D Ising model as a
function of reduced temperature $r$ and applied field $h$.}
\label{fig:mrh}
\end{figure}

\begin{figure}[thb]
\vspace{-0.4in}
\begin{center}
\epsfig{file=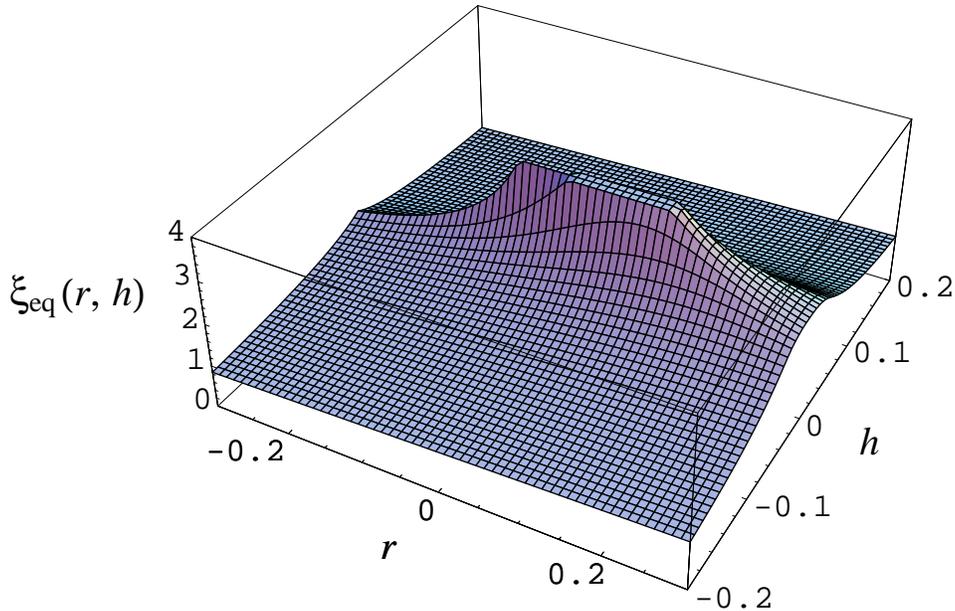,width=5.in}
\end{center}
\vspace{-0.5in}
\caption{Equilibrium correlation length as a function of reduced
temperature $r$ and applied field $h$.}
\label{fig:xieqrh}
\end{figure}
We want $\xi_{\rm eq}(r,h)$.  With $\xi_{\rm eq}(r,M)$ in hand,
we must now
obtain the magnetization $M(r,h)$. The most convenient form
for our purposes is the 
parametric equation of state (see
\cite{GuidaZinn, Brez,Wallace}):
\begin{equation}
\left \{
\begin{array}{ccl}
M & = & M_0 R^\beta \theta \\
h & = & h_0 R^{\beta \delta} \tilde{h}(\theta) = h_0 R^{\beta \delta} (\theta
- 0.76201 \, \theta^3 + 0.00804 \, \theta^5) \\
r & = & R (1 - \theta^2)
\end{array}
\right .
\label{eos-param}
\end{equation}
Here $M$, $r$ and $h$ are parametrized in terms of the ``radius''
$R\geq 0$ and ``polar angle'' $\theta$.  $\theta=0$ corresponds to
$r>0$, $h=0$; $\theta = \pm 1$ corresponds to $r=0$ with positive and
negative $h$ respectively. $\tilde{h}(\theta)$ is zero at $\theta=\pm 1.154$,
which corresponds to $r<0, h=\pm 0$.  The function $\tilde{h}(\theta)$ in
(\ref{eos-param}) is from Ref. \cite{GuidaZinn}, where it is
constructed to be consistent with all that is known from the
$\ep$-expansion, from perturbation theory, and from resummations
thereof.  We choose the normalization constants $M_0$ and $h_0$ so
that $M(r=-1, h=0)=1$ and $M(r=0, h=1)=1$.  This guarantees that along
the negative $r$ axis $M(r, h=0) = |r|^{-\beta}$ and
along the $h$ axis $M(r=0, h) = \mbox{sign}(h) |h|^{- \delta}$.
Numerically solving the last two equations of (\ref{eos-param}) for
$R$ and $\theta$ in terms of $r$ and $h$, we use the first one to
compute $M(r, h)$, shown in Figure 4.  
This allows us to obtain $\xi_{\rm eq}(r,h)$,
shown in Figure 5.


\begin{thebibliography}{99}

\bibitem{SRS1}
M. Stephanov, K. Rajagopal and E. Shuryak,
Phys. Rev. Lett. {\bf 81} (1998) 4816.

\bibitem{SRS2}
M. Stephanov, K. Rajagopal and E. Shuryak,
Phys. Rev. {\bf D60} (1999) 114028.

\bibitem{Review} 
For a recent review, see K. Rajagopal, to appear 
in Proceedings of Quark Matter '99, {\tt hep-ph/9908360}.

\bibitem{piswil}
R. Pisarski and F. Wilczek, Phys. Rev. {\bf D29} (1984) 338.

\bibitem{latticereview}
For reviews, see E. Laermann Nucl. Phys. Proc. Suppl. {\bf 63} (1998) 114
and A. Ukawa, Nucl. Phys. Proc. Suppl. {\bf 53} 
(1997) 106.

\bibitem{latticeTc} For example, S. Gottlieb et al., 
Phys. Rev. {\bf D55} (1997) 6852 and R.~Mawhinney, talk
at ISMD99, Providence, RI, 1999.  

\bibitem{NJL}  
A. Barducci, R. Casalbuoni, S. De Curtis, R. Gatto, G. Pettini, 
Phys. Lett. {\bf B231} (1989) 463;
S.P. Klevansky, Rev. Mod. Phys. {\bf 64} (1992) 649; 
A. Barducci, R.~Casalbuoni, G. Pettini and R. Gatto, Phys. Rev. {\bf D49}
(1994) 426.

\bibitem{steph} 
        M. Stephanov, Phys. Rev. Lett. {\bf 76} (1996) 4472;
        Nucl. Phys. Proc. Suppl. {\bf 53} (1997) 469.

\bibitem{ARW1}
M. Alford, K. Rajagopal and F. Wilczek, Phys. Lett. {\bf B422} (1998) 247.

\bibitem{RappETC}
R. Rapp, T. Sch\"afer,
E. V. Shuryak and M. Velkovsky, Phys. Rev. Lett. {\bf 81} (1998) 53.

\bibitem{bergesraj} 
J. Berges and K. Rajagopal, Nucl. Phys. {\bf B538} (1999)
215.

\bibitem{stephetal}
M. A. Halasz, A. D. Jackson, R. E. Shrock, M. A. Stephanov
and J.~J.~M.~Verbaarschot, Phys. Rev. {\bf D58} (1998) 096007.

\bibitem{PisarskiRischke1OPT}
R. Pisarski and D. Rischke, Phys. Rev. Lett. {\bf 83} 
(1999) 37.

\bibitem{CarterDiakonov}
G. Carter and D. Diakonov, Phys. Rev. {\bf D60} (1999) 016004.


\bibitem{rajwil}
F. Wilczek, Int. J. Mod. Phys. {\bf A7} (1992) 3911;
K. Rajagopal and F.~Wilczek, Nucl. Phys. {\bf B399} (1993) 395.

\bibitem{columbia}
F. Brown {\it et al}, Phys. Rev. Lett. {\bf 65} (1990) 2491.

\bibitem{kanaya}
JLQCD Collaboration, Nucl. Phys. Proc. Suppl. {\bf 73} (1999) 459.

\bibitem{oldkanaya}
Y. Iwasaki {\it et al}, Phys. Rev. {\bf D54} (1996) 7010.

\bibitem{PBM}
See, e.g., P. Braun-Munziger and J. Stachel,
Nucl. Phys. {\bf A606} (1996) 320.

\bibitem{NA49}
NA49 Collaboration, {\tt hep-ex/9904014}.

\bibitem{Mrow} St. Mr\'owczy\'nski, Phys. Lett. {\bf B430} (1998) 9.



\bibitem{laterfreezeout} The fact that larger systems freezeout
later has been established experimentally by seeing the $A$-dependence
of the freeze-out temperature via analyses of flow \cite{HS}, Coulomb
effects (H. W. Barz, J.~P.~Bondorf, J. J. Gaardhoje and H. Heiselberg,
Phys. Rev. {\bf C57} (1998) 2536), 
and pion interferometry (U. Heinz, Proceedings of Quark
Matter '97, {\tt nucl-th/9801050}).

\bibitem{HS} C. M. Hung and E. Shuryak, Phys. Rev. {\bf C57} (1998) 1891.

\bibitem{Heinz2}
B. Tomasik, U. A. Wiedemann and U. Heinz, {\tt nucl-th/9907096}; and
U. Heinz, private communication.

\bibitem{HoHa} 
P. C. Hohenberg and B. I. Halperin, Rev. Mod. Phys. {\bf 49} (1977) 435.

\bibitem{GuidaZinn}
R. Guida and J. Zinn-Justin, {\tt hep-th/9610223}; {\tt cond-mat/9803240}; 
and J. Zinn-Justin {\tt hep-th/9810193}.

\bibitem{Heinz}
E. Schnedermann and U. Heinz, Phys. Rev. {\bf C47} (1993) 1738; and 
{\bf C50} (1994) 1675; and U. Heinz, private communication.

\bibitem{OldShuryak}
E. Shuryak, Sov. J. Nucl. Phys. {\bf 16} (1973) 220.

\bibitem{Bravina}
L. V. Bravina {\it et al},  Phys. Rev. {\bf C60} (1999) 024904.

\bibitem{Tsypin}
See, e.g., K. Rummukainen, M. Tsypin, K. Kajantie, M. Laine and 
M.~Shaposhnikov,
Nucl. Phys. {\bf B532} (1998) 283.

\bibitem{Misha}
M. Stephanov, private communication.

\bibitem{Brez}
E.~Br\'ezin, J.~C.~Le Guillou, and J.~Zinn-Justin, in Phase
Transitions and Critical Phenomena {\bf 6}, (Academic Press, 1976),
ed. C.~Domb and M.~S.~Green.


\bibitem{Wallace}
D. Wallace in Phase
Transitions and Critical Phenomena {\bf 6}, (Academic Press, 1976),
ed. C.~Domb and M.~S.~Green.


\end{thebibliography}
\end{document}